# Bronchovascular Tree-Guided Weakly Supervised Learning Method for Pulmonary Segment Segmentation

Ruijie Zhao[1], Zuopeng Tan [2], Xiao Xue[2], Longfei Zhao[2], Bing Li[2], Zicheng Liao[1], Ying Ming[1], Jiaru Wang[1], Ran Xiao1, Sirong Piao[1], Rui Zhao[1], Qiqi Xu[2], Wei Song[1]

1.Department of Radiology, Peking Union Medical College Hospital, Chinese Academy of Medical Sciences and Peking Union Medical College, Beijing, China
2. Canon Medical Systems (China), Beijing, China

*Abstract*—Pulmonary segment segmentation is crucial for cancer localization and surgical planning. However, the pixel-wise annotation of pulmonary segments is laborious, as the boundaries between segments are indistinguishable in medical images. To this end, we propose a weakly supervised learning (WSL) method, termed Anatomy-Hierarchy Supervised Learning (AHSL), which consults the precise clinical anatomical definition of pulmonary segments to perform pulmonary segment segmentation. Since pulmonary segments reside within the lobes and are determined by the bronchovascular tree, i.e., artery, airway and vein, the design of the loss function is founded on two principles. First, segment-level labels are utilized to directly supervise the output of the pulmonary segments, ensuring that they accurately encompass the appropriate bronchovascular tree. Second, lobe-level supervision indirectly oversees the pulmonary segment, ensuring their inclusion within the corresponding lobe. Besides, we introduce a two-stage segmentation strategy that incorporates bronchovascular priori information. Furthermore, a consistency loss is proposed to enhance the smoothness of segment boundaries, along with an evaluation metric designed to measure the smoothness of pulmonary segment boundaries. Visual inspection and evaluation metrics from experiments conducted on a private dataset demonstrate the effectiveness of our method.

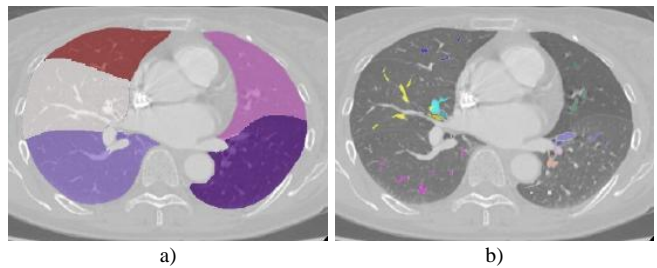

Fig. 1. Illustration of lung anatomy, a) annotation of lobe, b) annotation of bronchovascular tree. Regions annotated with differnent colors represent the charateristic regions belong to the different segments.

TABLE I.     Sub-structure of Lung Anatomy

| Lung | Lobe | Segment (LS for left segment, RS for right segment) |
|---|---|---|
| Left Lung | Left Upper Lobe | LS1/2, LS3, LS4, LS5 |
| | Left Lower Lobe | LS6, LS7/8, LS9, LS10 |
| Right Lung | Right Upper Lobe | RS1, RS2, RS3 |
| | Right Middle Lobe | RS4, RS5 |
| | Right Lower Lobe | RS6, RS7, RS8, RS9, RS10 |

## I. Introduction

For lung disease diagnosis, the computed tomography (CT) is often used in clinical treatment planning. The segmentation of lung sub-structure (e.g. lobes, segments) from CT images is benefit to the cancer locating and the surgical planning of resection. As shown in TABLE I. , human lungs are divided into 5 lobes and further divided into 18 segments totally, which is a hierarchical structure[1][[2].

The physical boundaries between the lobes are defined by the lobar fissures, which can typically be distinguished from CT images [1]. As shown in Fig. 1 a), the right lung includes 3 lobes, the upper lobe, the middle lobe, and the lower lobe and the left lung includes 2 lobes. Clear boundaries and lobar delineation make labeling of the lobes feasible, which drives the learning-based lobe segmentation [6][7][18][23] to achieve promising results.

However, the absence of distinct boundaries between the pulmonary segments within a lung lobe complicates the labeling of pulmonary segments, thereby hindering the advancement of fully supervised learning (FSL) methods for pulmonary segment segmentation. Fortunately, the bronchovascular tree, excluding intersegmental veins, within the lungs can be classified into distinct segment-level categories, as illustrated in Fig. 1 b). A viable way for pulmonary segment segmentation is to leverage the bronchovascular tree as guiding frameworks for the model.

A clear definition of pulmonary segments should satisfy three criteria [24][25]: first, the volume of a certain segment should be completely enclosed within the corresponding pulmonary lobe. Second, a certain segment should encompass the associated pulmonary artery and airway. Finally, the intersegmental veins serve as boundaries of two segments and the intrasegmental veins should be included by segments. In this paper, we propose an Anatomy-Hierarchy Supervised Learning (AHSL) method. AHSL adopts definition of pulmonary segments and utilizes lobe-level and segment-level labels to ensure that pulmonary segments are contained within the corresponding pulmonary lobe while also encompassing their associated bronchovascular tree.

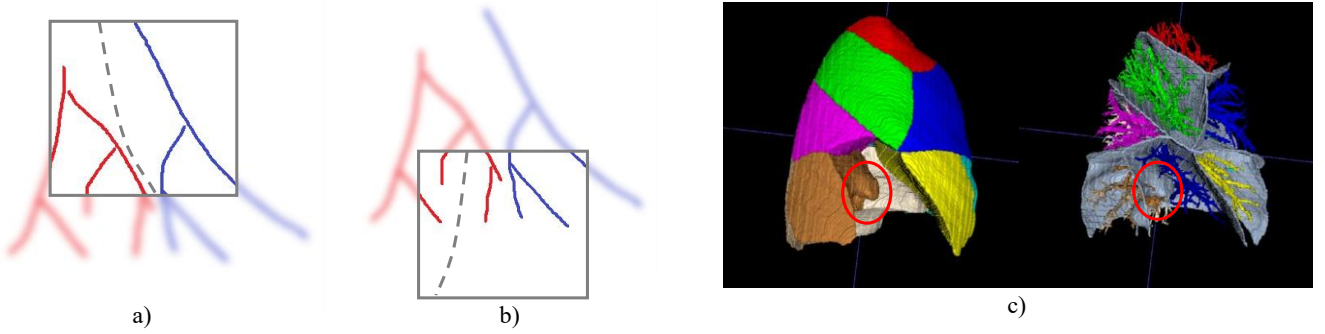

Fig. 2. An illustration of segmentation errors caused by the absence of bronchovascular structure. The red and blue skeletons in a) and b) represent the arterial skeletons of different pulmonary segments, while the gray boxes indicate the windows utilized during patch inference. The gray dashed lines represent the predicted boundaries of the pulmonary segments. a) Adequate bronchovascular structure results in accurate prediction of pulmonary segment boundary. b) Loss of structural information leads to incorrect prediction of pulmonary segment boundary. c) The segmentation result, with the red circle highlighting the errors in segmentation attributed to the absence of structural information.

AHSL primarily relies on the structure of the bronchovascular tree for pulmonary segment segmentation. However, when applying patch inference commonly used in 3D medical segmentation, structural integrity of bronchovascular tree is compromised. As illustrated in Fig. 2 b), in instances where two branches from different pulmonary segments are in close proximity, and the inference patch fails to encompass their parent branches, the model is likely to generate incorrect and discontinuous segment boundaries. To address this challenge, we propose a two-stage segmentation method. In the initial stage, the pulmonary artery and airway are segmented. The segmentation result then offers clear guidance for the second stage, thereby helping to get correct pulmonary segment boundaries. Besides, to achieve smooth boundaries in pulmonary segments, we propose a consistency loss that encourages the model to generate probability maps with locally consistent values, thereby producing smooth pulmonary segment masks. Finally, to quantitatively evaluate the smoothness of the segmentation result, we introduce a smoothness metric, denoted as $\#holes$, which measures the number of holes present in the axial, sagittal, and coronal planes.

The main contribution of this paper can be summarized to following items:

- We propose a novel Anatomy-Hierarchy Supervised Learning (AHSL) method, which utilizes pulmonary lobe and bronchovascular tree to guide pulmonary segment segmentation.
- We design a two-stage segmentation strategy to provide priori knowledge of bronchovascular for pulmonary segment segmentation.
- We propose a consistency loss to enhance the smoothness of the pulmonary segment boundaries.
- A smoothness evaluation metric, $\#holes$, is proposed to quantify the smoothness of the segmentation results.

## II. RELATED WORK

For segmenting lung and related structures from images automatically, there are several works that have been purposed by researchers from different kinds of domains [[2]]. Recent years, deep learning based methods become the state-of-the-art in numerous medical image segmentation tasks. The technique has been widely used for the pixelwise segmentation tasks such as lung [3][4], lobe [5][10][18], airway[11][12], and vessel[13][14]. These methods focus on different kinds of difficulties (such as the complicated disease patterns, the tiny number of annotated data, the constrained model size) and obtain promising results.

Recently, several methods for pulmonary segment segmentation have been proposed. Kuang et al. [16] introduce an implicit-function-based model that enables the reconstruction of pulmonary segments by querying the class of spatial coordinates. Additionally, Koh et al. [19] employ a probabilistic approach utilizing Gaussian Mixture Models to capture the anatomical relationships of pulmonary segments and achieve effective segmentation. These methods belong to FSL method, which means the pixelwise annotation of segmentation target is required for model training. However, the annotation of pulmonary segment is relatively difficult because of the invisible boundary.

In contrast to FSL, weakly supervised learning (WSL) offers a pragmatic solution. Bai et al. [20] utilize the internal central box of the pulmonary segment as weak labels for pulmonary segment segmentation. Recently, IPGN [21] achieves pulmonary mask labeling by extracting point and graph features from bronchovascular tree. These two WSL methods respectively employ segment-level pseudo labels and bronchovascular tree features to achieve pulmonary segment segmentation. However, they both overlook the actual clinical anatomical definition of pulmonary segments, resulting in pulmonary segments that do not adequately encompass the corresponding bronchovascular structures.

## III. METHOD

### A. Segmentation of bronchovascular tree

To avoid the loss of structural information in bronchovascular tree caused by patch inference, it is common practice to either increase the patch size or decrease the image resolution. However, enlarging the patch size results in higher memory consumption, while reducing the image resolution leads to blurriness and a loss of structural details. To this end, we directly segment the bronchovascular tree in the first stage, identifying the categories for each segment to facilitate the subsequent segmentation of pulmonary segment. Note that intersegmental veins exist at the boundaries of pulmonary segments, making precise categorization challenging. Consequently, we do not segment veins in the first stage.

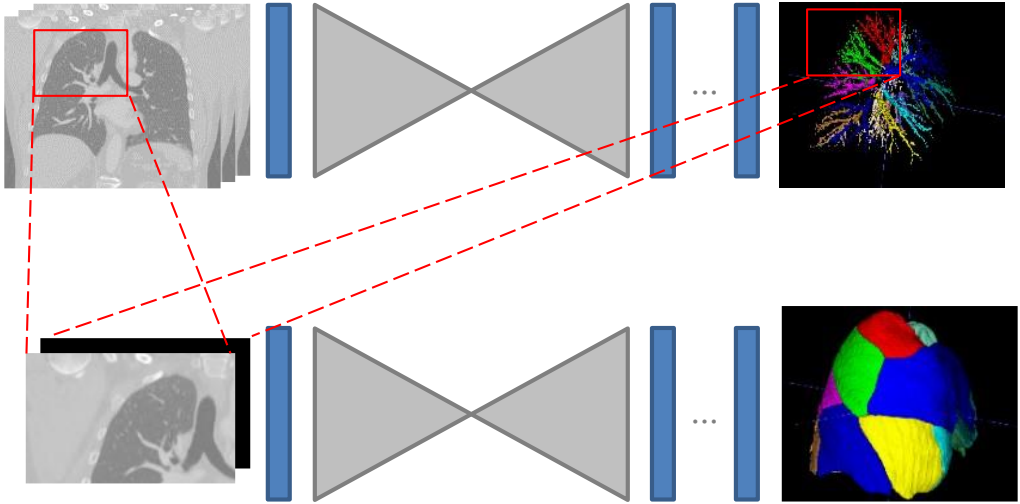

Fig. 3. The overall framework of AHSL. In the first stage, the model performs the segmentation of the artery and airway. Subsequently, the segmentation results from the first stage are concatenated with the original image and fed into the second-stage model, which then performs the pulmonary segment segmentation.

We select U-Net [14] as the backbone network. The parameters of U-Net adhere to default setting of nnUNet [15], which is known as the framework for fast and effective segmentation. Giving an image, the output of the network is a probability map comprising 19 channels. These channels represent 19 classes, which include the background and 18 segments corresponding to the artery and airway. The bronchovascular structures, i.e., artery and airway, are easy to label. Therefore, we employ FSL to train the model in the first stage. Specifically, we use the common Dice and cross entropy as the loss functions for the first stage. Once getting the bronchovascular structure mask, we input the image and the mask into the second stage of the model. The bronchovascular structure mask provides more explicit a priori information, which encourages the model to accurately identify the boundaries of pulmonary segments. This is particularly beneficial when the input image patches consist solely of monotone bronchovascular structures. Anatomy-Hierarchy Supervised Learning

As shown in Fig. 3, we concatenate the bronchovascular tree with the image and input them into the second stage of the network. The network structure of the second stage mirrors that of the first stage, with the exception that the input channel of the first layer is set to 2.

Due to the indistinguishable boundaries of pulmonary segments, the annotation of pulmonary segments is difficult. Abandoning the pulmonary segment mask, we utilize the bronchovascular tree mask and the pulmonary lobe mask to guide the learning process of model. As shown in Fig. 3, the bronchovascular tree mask comprises 18 segment-level categories, referred to as segment-level labels. In contrast, the pulmonary lobe mask consists of 5 lobe-level categories, known as lobe-level labels. The lobe-level labels are employed to encompass the corresponding output pulmonary segments, thereby providing indirect supervision to the segmentation model. Furthermore, these lobe-level labels serve to mitigate the risk of over-segmentation within the pulmonary segments. In a pulmonary lobe, segment-level labels prompt the output pulmonary segments to include their corresponding bronchovascular tree, which can further divide the pulmonary lobe into detailed pulmonary segments. It satisfies the clinical definition of pulmonary segment. In summary, this anatomy-hierarchy supervised learning keeps

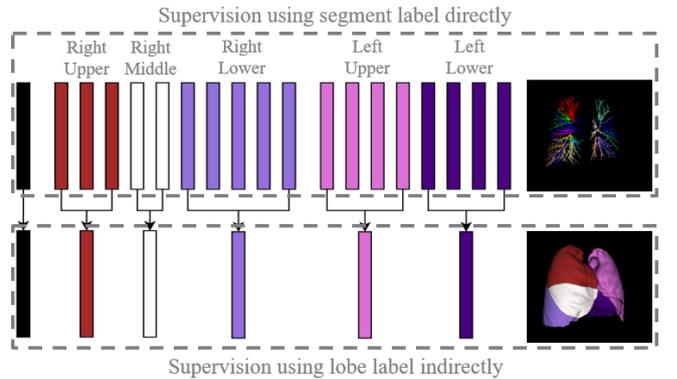

Fig. 4. Illustration of the supervision. The upper section of the figure presents 19 probability map channels corresponding to the background and 18 segments produced by the network. The lower section displays 6 background channels and 5 lobes. The probability map channel for a lobe can be derived from the probability map channels of the segments contained within that lobe.

TABLE II. NOTATION OF REGIONS

| Notation | Description |
| --- | --- |
| B | The set of voxels belongs to background region |
| BV | The set of voxels belongs to bronchovascular region |
| L | The set of voxels belongs to lobe region |

the overall structure of lung while ensuring a accurate delineation of the pulmonary segments.

Specifically, segment-level and lobe-level labels divide regions of an image into three categories, i.e., background region (B), bronchovascular region (BV), and lobe region (L), as outlined in TABLE II. . The format of the supervision is shown in Fig. 4. The design of the loss function is based on the following two fundamental principles,

- Segment-level label directly supervises the voxels located in the BV, while voxels around BV are

- encouraged to incorporate as many as possible into their corresponding BV.
- Lobe-level label indirectly supervises the voxels located in L. It means the voxels should be one of several segments within this lobe, e.g. the voxel locates in right Middle lobe, it should belong to RS4 or RS5.

According to the two principles, the loss function of directly supervised part is as follows:

$$Loss_{directly} = Recall_{BV} + CE_{BV}, \quad (1)$$

where $Recall_{BV}$ promotes the inclusion of voxels surrounding the BV within the BV itself. And $CE_{BV}$ directly guarantees that the voxels contained within the BV are accurately classified.

The loss function of indirectly supervised part is the sum of dice loss and CE loss on voxel set L,

$$Loss_{indirectly} = Dice_L + CE_L. \quad (2)$$

The probability that a voxel belongs to a specific lobe is determined by the maximum probability of that voxel belonging to any of the segments contained within that lobe, as illustrated in Fig. 5.

### B. Boundary consistency

Although AHSL can effectively segment pulmonary segments within the pulmonary lobes, the boundaries of these segments are often irregular due to insufficient consistency constraints. To this end, we propose a consistency loss that enforces the probabilities of a six-neighborhood at any position within the output probability map to be closely aligned. This approach enhances the smoothness of the segmentation results, particularly at the boundaries of the pulmonary segments.

The consistency loss is the L1-norm of Laplacian of the probability map,

$$Loss_{consistence} = ||\nabla \cdot \nabla (probability\ map)||_1. \quad (3)$$

For example, the value of voxel $(z, y, x)$ of channel c is $p_{c,z,y,x}$, the Laplacian is:

$$\nabla \cdot \nabla p_{c,z,y,x} = p_{c,z+1,y,x} + p_{c,z-1,y,x} + p_{c,z,y+1,x} + p_{c,z,y-1,x} + p_{c,z,y,x+1} + p_{c,z,y,x-1} - 6p_{c,z,y,x}. \quad (4)$$

It measures the difference of the probability of center voxel with neighbor voxels, and it benefits the boundary smoothness.

Overall, the formula of loss function is:

$$Loss = Loss_{directly} + \lambda_1 Loss_{indirectly} + \lambda_2 Loss_{consistence}, \quad (5)$$

where $\lambda_1, \lambda_2$ are hyperparameters. They are both set as 1 in our experiment.

## IV. EXPERIMENTS

### A. Dataset

We retrospectively collect chest CT images with layer thickness between 0.5 mm and 1.25 mm, including both contrast-enhanced and non-contrast CT. The dataset comprises images from normal subjects as well as individuals with pulmonary diseases such as nodules and tumors. We exclude images presenting severe conditions that hindered accurate segmentation, including severe lesions that could not be judiciously segmented, bronchial occlusions without vascular signals due to tumor compression, and bronchial mucus plugs that obstructed airway segmentation.

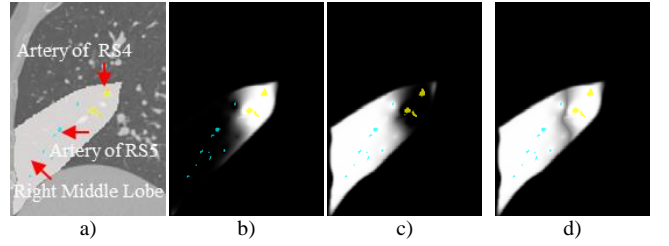

Fig. 5. Illustration of the indirect supervision. The annotation of artery of RS4 (yellow) and RS5 (blue) are shown in all sub-figures for better visualization, a) the annotation of region of right middle lobe, arteries of RS4 and RS5, b) the optimized probability map of RS4, c) optimized probability map of RS5, d) the maximum of probabilitity map of RS4 and RS5, which is used for indirect loss.

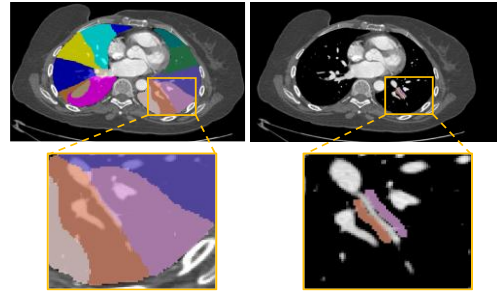

Fig. 6. The annotation of intersegmental vein. The left figure shows that the vein is situated at the boundary between two polmonary segments. Consequently, two sides of vein are labeled with scribble.

We finally collect 200 CT data, including 104 Computed Tomography Pulmonary Angiography (CTPA) and 96 Non-Contrast Computed Tomography (NCCT). We use 159 data (85 CTPA and 74 NCCT) for training and 41 data (19 CTPA and 22 NCCT) for testing. The pulmonary lobes and bronchovascular structures are annotated by one radiologist with 10 years of experience. Note that intersegmental veins are located at the borders of two pulmonary segments and could not be classified into distinct segment categories, as illustrated in Fig. 6. Therefore, we adopt scribble annotation to label two sides of the vein as their corresponding categories.

### B. Implement Details

For comparing proposed method with the FSL method, the ground truth (GT) of segments are synthesized using the discrete Euclidean distance to the bronchovascular structures described in [3]. And the compared model is trained using same network structure and the loss function of $Loss := Dice + CE$.

*1) Preprocessing:* The data are re-sampled to spacing 0.8mm×0.73mm×0.73mm and normalized according to the mean and standard deviation. The input patch size is 128×192×192.

*2) Data Augmentation:* Rescaling of grey value and rotation augmentation is used during training time. Note that the test time augmentation (TTA) is not used.

*Optimization:* Both two stages are trained using Stochastic Gradient Descent (SGD) with initial learning rate of 0.01 and batch size of 2. In each epoch, 250 batches are random cropped from augmented images and used for optimizing the network parameters, and 1000 epochs are executed totally.

TABLE III. QUANTITATIVE RESULTS

| Modality | Mthod | $Dice_{Artery}\uparrow$ | $Dice_{Airway}\uparrow$ | #holes↓ |
|---|---|---|---|---|
| CTPA | IPGN [21] | 0.702±0.241 | 0.512±0.318 | 182.80±142.49 |
|  | AHSL | **0.923±0.143** | **0.924±0.157** | **1.47±2.43** |
| NCCT | IPGN [21] | 0.713±0.259 | 0.551±0.317 | 209.83±112.35 |
|  | AHSL | **0.933±0.160** | **0.934±0.156** | **1.17±2.25** |

TABLE IV. ABLATION STUDY

| Modality | Two stage strategy | Consistence Loss | $Dice_{Artery}\uparrow$ | $Dice_{Airway}\uparrow$ | #holes↓ |
|---|---|---|---|---|---|
| CTPA |  |  | 0.906±0.170 | 0.921±0.169 | 36.15±44.52 |
|  | ✓ |  | 0.922±0.149 | 0.923±0.172 | 21.14±26.35 |
|  | ✓ | ✓ | **0.923±0.143** | **0.924±0.157** | **1.47±2.43** |
| NCCT |  |  | 0.920±0.164 | 0.928±0.174 | 40.01±52.21 |
|  | ✓ |  | 0.929±0.167 | **0.934±0.165** | 20.84±31.91 |
|  | ✓ | ✓ | **0.933±0.160** | 0.933±0.156 | **1.17±2.25** |

### C. Evaluation Metrics

Following [[15]], the Dice score of the Mapped Artery ($Dice_{Artery}$) and the Mapped Airway ($Dice_{Airway}$) is used for evaluating whether the bronchovascular structures are included in the correct segments. Specifically, artery and airway masks are labeled according to the pulmonary segment in which they are situated. Subsequently, the Dice score is computed between the labeled mask and the ground truth.

To evaluate the smoothness of the pulmonary segment boundaries, we propose a smoothness evaluation metric called #holes . As shown in Fig. 7, this metric separately calculates the number of holes in axial, sagittal, and coronal slices, subsequently summing these values. #holes represents topological complexity that is inversely proportional to geometric smoothness. This aligns with computational principles, where a reduced number of enclosed regions signifies well-defined and anatomically consistent segment boundaries.

### D. Results

#### 1) Comparision with Baseline Method

The quantitative results are shown in TABLE III. The proposed AHSL achieves over 30% improvement than IPGN for $Dice_{Artery}$ on both NCCT and CTPA data. For $Dice_{Airway}$, AHSL shows a 80% improvement on NCCT and a 70% improvement on CTPA. The IPGN utilizes the point and graph features of a singular bronchovascular tree for bronchovascular labeling, subsequently employing the learned features for pulmonary segment segmentation. However, due to the singularity of the bronchovascular tree features and the absence of supervision from other voxels within the pulmonary segment, the segmentation of pulmonary tissue outside the bronchovascular tree is suboptimal.

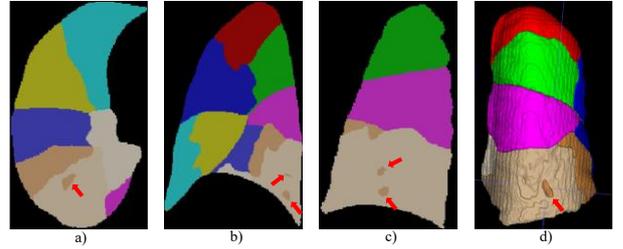

Fig. 7. Example of #holes. The red arrow highlights the hole on (a) axial, (b) sagittal, (c) coronal, and (d) 3D mask.

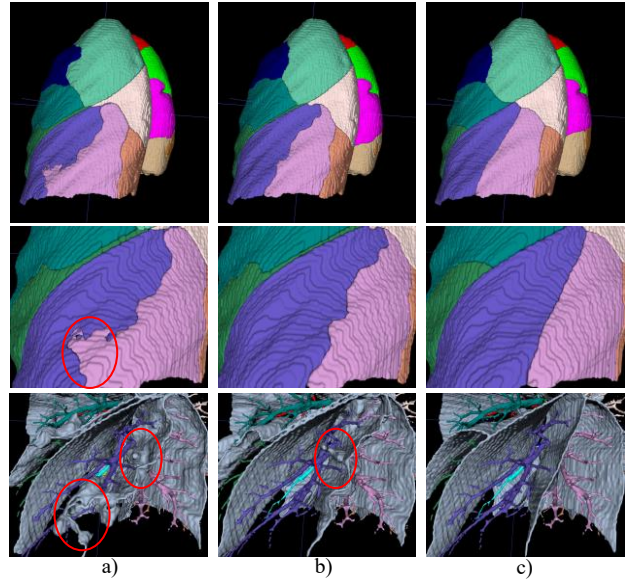

Fig. 8. Comparison of results with different components. The red circles in the figure highlight areas of incorrect segmentation or regions that lack smoothness. a) Results after removing the two-stage strategy and consistency loss. b) Results after omitting the consistence loss. c) Results with all components.

#### 2) Ablation Study

As demonstrated in TABLE IV. , we conduct three experiments to validate the efficacy of our proposed component. We utilize a simplified version of ASHL, which removes the first stage and the consistency loss. By incorporating the artery and airway segmentation model in the first stage, the model better finds the cues within the bronchovascular structure, resulting in improvements of 1.8% (CTPA), 1.0% (NCCT) in $Dice_{Artery}$. Besides, $\#holes$ is improved by 42% on CTPA and 48% on NCCT. The introduction of consistency loss restrict consistence of neighbor voxels and enables AHSL to get smoother pulmonary segment boundaries. As a result, with $Dice_{Artery}$ and $Dice_{Airway}$ maintained, $\#holes$ is improved by 93% on CTPA and 94% on NCCT, respectively.

TABLE V. SUBJECTIVE SCORES BY TWO RADIOLOGISTS ON EXTERNAL DATA

| Case | 1 | 2 | 3 | 4 | 5 | 6 | 7 | 8 | 9 | 10 | 11 | 12 | 13 | 14 | 15 | 16 | 17 | 18 | 19 | 20 | 21 | Mean±Std |
|---|---|---|---|---|---|---|---|---|---|---|---|---|---|---|---|---|---|---|---|---|---|---|
| Doctor_1 | 4 | 3 | 3 | 4 | 4 | 4 | 4 | 3 | 2 | 3 | 3 | 4 | 4 | 4 | 4 | 4 | 4 | 4 | 4 | 4 | 4 | 3.67±0.56 |
| Doctor_2 | 4 | 3 | 3 | 4 | 4 | 4 | 4 | 4 | 2 | 3 | 3 | 4 | 4 | 4 | 4 | 4 | 4 | 4 | 4 | 4 | 4 | 3.71±0.55 |

As illustrated in Fig. 8, the visualization results of the ablation experiment are presented. In Fig. 8 a), it is evident that the boundaries of the pulmonary segments display significant irregularities in the areas highlighted in red in the second and third rows. Furthermore, the pink pulmonary segment region inaccurately encompasses the purple arteries, which can be attributed to the deficiency in bronchovascular structural information. The incorporation of bronchovascular prior knowledge into the two-stage strategy lead to an improvement in segmentation errors, as demonstrated in Fig. 8 b). Although segmentation accuracy is improved, the pulmonary segment boundaries still exhibit some irregularities. However, after the introduction of consistency loss, the issue of irregular boundaries was completely resolved as shown in Fig. 8 c).

*3) Validation on External data*

To further validate the robustness and clinical validity of the AHSL, we conduct an external validation study using 21 multi-center datasets acquired from three different equipment. Two experienced radiologists independently evaluate the segmentation results of AHSL through a standardized scoring system. The five-point scale was defined as follows: 1 (segmentation failed, results deemed unacceptable), 2 (partially inaccurate with missing segments, results unreliable), 3 (generally accurate with minor boundary imperfections, results acceptable), 4 (accurate segmentation with uneven boundaries, results satisfactory), and 5 (excellent accuracy with smooth segment boundaries, results perfect). As demonstrated in Table 5, AHSL achieved mean scores of 3.67±0.56 and 3.71±0.55 across the two independent evaluations. These findings corroborate the technical robustness and clinical applicability of AHSL in pulmonary segment segmentation.

## V. CONCLUSION

In this study, a novel anatomy-hierarchy supervised learning (AHSL) method is proposed for pulmonary segment segmentation. The AHSL is guided by the bronchovascular tree and adheres strictly to the definition of pulmonary segments for segmentation purposes. Furthermore, the two-stage segmentation strategy effectively extracts prior information from the bronchovascular tree, enabling accurate pulmonary segment segmentation even when bronchovascular structure is incomplete. And we introduce a consistency loss that enhances the smoothness of the pulmonary segment boundaries, thereby making the segmentation results more applicable for clinical use. Lastly, a smoothness metric, $\#holes$, is proposed to measure the smoothness of the pulmonary boundaries.

REFERENCES

[1] van Rikxoort, E., et al.: Automatic Segmentation of Pulmonary Segments From Volumetric Chest CT Scans. IEEE transactions on medical imaging (2009).

[2] Stoecker, C., et al.: Determination of lung segments in computed tomography images using the Euclidean distance to the pulmonary artery. Medical physics (2013).

[3] van Rikxoort, E., et al.: Automated segmentation of pulmonary structures in thoracic computed tomography scans: a review. Phys Med Biol (2013).

[4] Harrison, A., et al.: Progressive and Multi-path Holistically Nested Neural Networks for Pathological Lung Segmentation from CT Images. International Conference on Medical Image Computing and Computer-Assisted Intervention (MICCAI) (2017).

[5] Hofmanninger, J., et al.: Automatic lung segmentation in routine imaging is primarily a data diversity problem, not a methodology problem. European Radiology Experimental (2020).

[6] Ferreira, F., et al.: End-to-End Supervised Lung Lobe Segmentation. International Joint Conference on Neural Networks (IJCNN) (2018).

[7] Wang, W., et al.: Automated Segmentation of Pulmonary Lobes using Coordination-Guided Deep Neural Networks. International Symposium on Biomedical Imaging (ISBI) (2019).

[8] Garcia-Uceda Juarez, A., et al.: Automatic airway segmentation from computed tomography using robust and efficient 3-D convolutional neural networks. Scientific Reports (2021).

[9] Nadeem, S., et al.: A CT-Based Automated Algorithm for Airway Segmentation Using Freeze-and-Grow Propagation and Deep Learning. IEEE transactions on medical imaging (2020).

[10] Cui, H., et al.: Pulmonary Vessel Segmentation based on Orthogonal Fused U-Net++ of Chest CT Images. International Conference on Medical Image Computing and Computer-Assisted Intervention (MICCAI) (2019).

[11] Tan, W., et al.: Automated vessel segmentation in lung CT and CTA images via deep neural networks. Journal of X-Ray Science and Technology (2021).

[12] Luo, X., et al.: Scribble-Supervised Medical Image Segmentation via Dual-Branch Network and Dynamically Mixed Pseudo Labels Supervision. International Conference on Medical Image Computing and Computer-Assisted Intervention (MICCAI) (2022).

[13] Liu, X., et al.: Weakly Supervised Segmentation of COVID19 Infection with Scribble Annotation on CT Images. Pattern Recognition (2021).

[14] Ronneberger, O., et al.: U-Net: Convolutional Networks for Biomedical Image Segmentation. International Conference on Medical Image Computing and Computer-Assisted Intervention (MICCAI) (2015).

[15] Isensee F., et al.: nnU-Net: a self-configuring method for deep learning-based biomedical image segmentation. Nature Methods (2021).

[16] Kuang, K., et al.: What Makes for Automatic Reconstruction of Pulmonary Segments. International Conference on Medical Image Computing and Computer-Assisted Intervention (MICCAI) (2022).

[17] Aktouf, Z., et al.: A 3d holes closing algorithm. Lecture Notes in Computer Science (1996).

[18] Su. Q., et al: Automatic Lobe Segmentation Using Attentive Cross Entropy and End-to-End Fissure Generation. International Symposium on Biomedical Imaging (ISBI) (2023).

[19] Koh, S., et al. PSGMM: Pulmonary Segment Segmentation Based on Gaussian Mixture Model. International Workshop on Shape in Medical Imaging. Cham: Springer Nature Switzerland, 2024: 18-32.

[20] Bai, Y., et al. Pulmonary Segments Segmentation with Hierarchical Weak Labels. International Symposium on Biomedical Imaging (ISBI), 2023: 1-5.

[21] Xie, K., et al. Efficient Anatomical Labeling of Pulmonary Tree Structures via Deep Point-graph Representation-based Implicit Fields. Medical image analysis, 2025, 99: 103367.

[22] W. Xie, C. Jacobs, J. -P. Charbonnier and B. van Ginneken, Relational Modeling for Robust and Efficient Pulmonary Lobe Segmentation in CT Scans, in *IEEE Transactions on Medical Imaging*, vol. 39, no. 8, pp. 2664-2675, Aug. 2020, doi: 10.1109/TMI.2020.2995108.

[23] Wang H, Wang X, Du Y, et al. Prediction of lymph node metastasis in papillary thyroid carcinoma using non-contrast CT-based radiomics and deep learning with thyroid lobe segmentation: A dual-center study. European Journal of Radiology Open, 2025, 14: 100639.

[24] Frick A E, Van Raemdonck D. Segmentectomies[J]. Shanghai Chest, 2017, 1(4).

[25] Oizumi H, Kato H, Endoh M, et al. Techniques to define segmental anatomy during segmentectomy[J]. Annals of cardiothoracic surgery, 2014, 3(2): 170.